\documentclass[aps,pre,superscriptaddress,showpacs,floatfix,english]{revtex4}
\usepackage[latin1]{inputenc}
\usepackage{color,psfig,graphics,graphicx,psfrag,amsmath,amssymb,amsfonts,rotating, babel, algorithm, algorithmic}

\newcommand{\sbf}[1]{{\sf{\textbf{#1}}}}
\newcommand{\RETURN}{\STATE \sbf{return} }
\newcommand{\INPUT}[1]{\tt\flushleft\sbf{Input} : #1\\}
\newcommand{\OUTPUT}[1]{\sbf{Output} : #1\vspace{\baselineskip}}
\newcommand{\GOTO}{\STATE \sbf{goto} line }

\makeatletter

\begin{document}
\title{Survey-propagation decimation through  distributed local computations}

\author{Jo\"el Chavas}
\email[]{chavas@isiosf.isi.it}
\affiliation{ISI Foundation, viale Settimo Severo 65, 10133 Torino, Italy.}

\author{Cyril Furtlehner}
\email[]{furtlehn@lptms.u-psud.fr}
\affiliation{LPTMS, b\^atiment 100, Universit\'e Paris-Sud, Centre scientifique d'Orsay, 15 rue Georges Cl\'emenceau, 91405 Orsay Cedex, France.}

\author{Marc M\'ezard}
\email[]{mezard@lptms.u-psud.fr}
\affiliation{LPTMS, b\^atiment 100, Universit\'e Paris-Sud, Centre scientifique d'Orsay, 15 rue Georges Cl\'emenceau, 91405 Orsay Cedex, France.}

\author{Riccardo Zecchina} 
\email[]{zecchina@ictp.trieste.it}
\affiliation{ICTP, Strada Costiera 11, I-34100 Trieste, Italy.}

\begin{abstract}
We discuss the implementation  of two {\it distributed} solvers of the random K-SAT problem,
based on some development of the recently introduced Survey Propagation (SP) algorithm. 
The first solver, called the ``SP diffusion algorithm'' diffuses as dynamical  information  the
maximum bias over the system, so that variables-nodes can decide to freeze in a self-organized
way, each variable taking its decision on the basis of purely local
information. 
The second solver, called the ``SP reinforcement  algorithm'', makes use of time-dependent 
external forcings on each variable, which are adapted in time
in such a way that the algorithm approaches its estimated closest solution. 
Both methods allow to find a solution of the random 3-SAT problem
 in a range of parameters comparable with the best previously described serialized
solvers. The simulated time of convergence towards a solution ( if these solvers were implemented
on a fully parallel device) grows as  $log(N)$.

\end{abstract}

\maketitle

\section{Introduction}

Recent progress in the statistical physics of disordered systems has provided
results of algorithmic interest:  the connection between the geometrical
structure of the space of solutions in random NP-complete constraint
satisfaction problems (CSP) and the onset of exponential regimes in search
algorithms \cite{TCS,MPZ} has been at least partially clarified. More
important, the analytical methods have been converted into a new class of
algorithms \cite{MZ} with very good performance \cite{BMZ}.

Similarly to statistical physics models, a generic  CSP is composed of many discrete variables
which interact through constraints, where each constraint   involves  a small number of variables.
When  CSPs  are extracted at random from non trivial ensembles there appear phase transitions 
as the constraint density (the ratio of constraints to variables) increases: for small densities 
the problems are satisfiable with high probability 
whereas when the density  is larger than a critical value  the problems become unsatisfiable with high probability.
Close to the phase boundary on the satisfiable side, most of the algorithms are known to take (typically) 
exponential time to find solutions.
The physical interpretation for the onset of such exponential regimes in random  hard  CSP consists in 
a trapping of local search  processes  in metastable states.
  Depending on the models and on the details of the
process the long time behaviour may be dominated by different types of metastable states.
However a  common feature which can be observed numerically is that for
simulation times which are sub-exponential in the size of the problem
there exists an extensive gap in the number of violated constraints 
which separates the blocking configurations from the optimal ones. 
 Such behavior can be tested on concrete random instances which therefore constitute 
a computational benchmark for more general algorithms.

In the last few years there has been a great progress in the  study of phase transitions 
in  random CSP which has produced new algorithmic tools:
 problems which were considered to be
algorithmically hard for local search algorithms, like for instance
random K-SAT close to a phase boundary, turned out to be efficiently
solved by the so called  Survey Propagation (SP) algorithm \cite{MZ}
arising from the replica symmetry broken (RSB) cavity approach to
CSP. According to the statistical physics analysis, close to
the phase transition point the solution space breaks up into many
smaller clusters \cite{MPZ,MZ}.  Solutions in separate clusters are
generally far apart. This picture has been confirmed by rigorous results, first  on
the simple case of the XORSAT problem \cite{cocco,MRZ}, and more recently for
the satisfiability problem \cite{MMZ,AR}.
Moreover,  the physics analysis indicates that clusters which correspond
to partial solutions ---which satisfy some but not all of the
constraints--- are exponentially more numerous than the clusters of
complete solutions and act as dynamical traps for local search
algorithms. SP turns out to be able to deal efficiently with the
proliferation of such clusters of metastable states.

The SP algorithm consists in a message-passing technique which is
tightly related to the better known Belief Propagation algorithm (BP)
\cite{BP} ---recently applied with striking success in the
decoding of error-correcting codes based on sparse graphs encodings~\cite{M,RicUrb,G, KFL, LTS, luby}---
but which differs from it in some crucial point. The messages sent
along the edges of the graph underlying the combinatorial problem
describe in a probabilistic way the cluster-to-cluster fluctuations of
the optimal assignment for a variable; while BP performs a ``white''
average over all the solutions, SP tells us which is the probability
of picking up a cluster at random and finding a given variable forced
to take a specific value within it (``frozen'' variable).  Once the
iterative equations have reached a fixed-point of such probability
distributions (called ``surveys'' because they capture somehow the
distribution of the expectation  in the different clusters), it becomes
possible in general to identify the variables which can be safely
fixed and to simplify the problem accordingly \cite{MZ,BMZ}.
This procedure, which   is intrinsically serial, is known as  the SP-inspired  decimation algorithm (SID).

From  the experimental point of view, the SID
has been efficiently used to solve many instances in the hard region of different
satisfiability problems and of the graph coloring problem, including instances too
large for any earlier method \cite{MZ}. For example, for random 3-SAT, instances
close to the threshold, up to sizes of order $10^7$ variables were solved and the
computational time in this regime was found experimentally to
scale roughly as $N (\log N)^2$.

In the present manuscript, we show  how the SP decimation algorithm can be made fully distributed.
This opens the possibility of  a parallel implementation which would lead to a further drastic
reduction in the computational cost. 

The paper is organized as follows: in part II we introduce some notation and the K-SAT
problem, in part III we recall the SP algorithm along with the serial decimation 
procedure, and parts IV  and V are devoted to explaining our new parallel solvers, respectively the first one 
based on information diffusion and the second one based on external forcing on the variables.

\section{The SAT problem and its factor graph representation}
\label{sect_fg}

Though the results we shall discuss are expected to hold for many types of CSP, here we consider
just one representative case : the K-SAT problem.

A  K-SAT formula consists of $N$ boolean variables $x_i \in \{0,1\}$ , $i \in \{ 1,\ldots,N \}$, with $M$ constraints, in which each
constraint is a clause, which is the logical OR ($\vee$) of the variables it contains or of
their negations.

Solving the K-SAT formula  means finding an assignment of the $x_i \in \{ 0,1\}=\{$directed,negated$\}$ which is such that all the $M$ clauses are true. In the literature, such an assignment can indifferently be called solution, satisfying assignment or ground state if one uses the statistical physics jargon.

In the case of randomly generated K-SAT formulas (variables appearing in the clauses chosen uniformly at random without repetitions and negated with probabiliy $1/2$)  a sophisticated phase transition phenomenon sets in :  when the number of constraints $M=\alpha N$ is small, the solutions of the K-SAT formulas are distributed close one to each other over the whole $N-$dimensional space, and the problem can easily be solved by the use of classical local search algorithms. 
Conversely, when $\alpha$ is included in a narrow region $\alpha_d<\alpha<\alpha_c$, the problem is still satisfiable but the now limited solution phase breaks down into an exponential number of clustered components. Solutions become grouped together into clusters which are fart apart one from the other. This range of parameters is known as the hard-SAT phase: classical local search algorithms get trapped by strictly positive energy configurations which are exponentially more numerous than the ground state ones.

A clause $a$ is characterized by the set of variables
$i_1,...,i_K$ which it contains, and the list of those which are
negated, which can be characterized by a set of $K$ numbers $J^a_{i_r}
\in \{ \pm 1\}$ as follows.  The clause is written as
\begin{equation}
 \left ( z_{i_1} \vee ...\vee z_{i_r} \vee ... \vee z_{i_K} \right)
\end{equation}
where $z_{i_r}=x_{i_r}$ if $J^a_{i_r}=-1$ and $z_{i_r}=\bar x_{i_r}$
if $J^a_{i_r}=1$ (note that a positive literal is represented by a
negative $J$). The problem is to find whether there exists an
assignment of the $x_i \in \{ 0,1\}$ which is such that all the $M$
clauses are true.  We define the total cost $C$ of a configuration
${\bf x}=(x_1,...,x_N)$ as the number of violated clauses.

\begin{figure}
\centering
\includegraphics[width=10.cm]{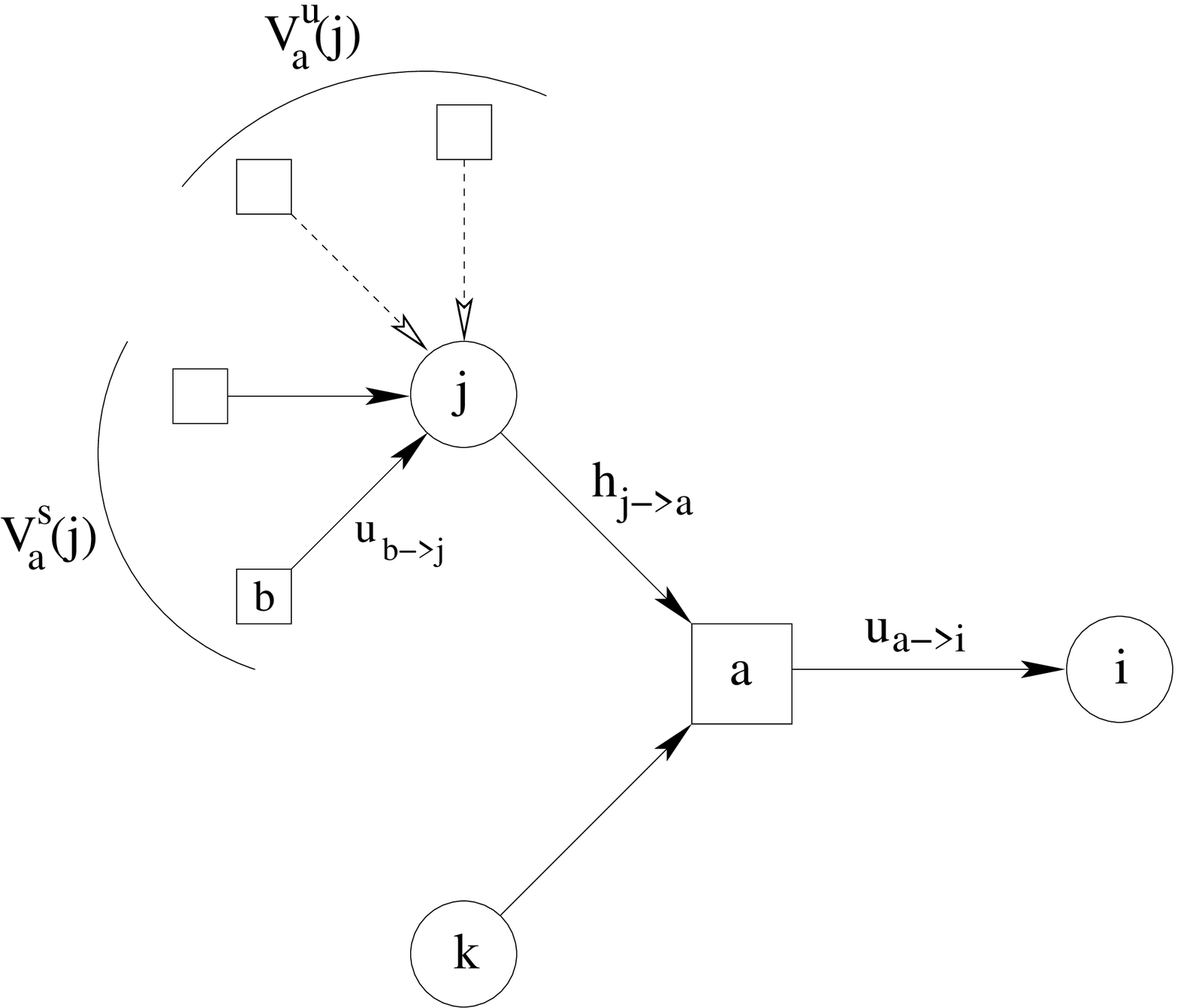}
\caption{ A function node $a$ and its neighborhood. The survey of
the cavity-bias $u_{a \to i}$ can be computed from the knowledge
of the joint probability distribution for all the cavity-biases in
the set $U$, i.e. those coming onto all variable nodes $j$
neighbors of $a$ (except $j=i$). } \label{fig_iter}
\end{figure}

In the so called "factor graph" representation, 
the SAT problem can be schemed as follows (see
Fig.~\ref{fig_iter}). Each of the $N$ variables is associated to a
vertex in the graph, called a ``variable node'' (circles in the
graphical representation), and each of the $M$ clauses is associated
to another type of vertex in the graph, called a ``function node''
(squares in the graphical representation). A function node $a$ is
connected to a variable node $i$ by an edge whenever the variable
$x_i$ (or its negation) appears in the clause $a$.  In the graphical
representation, we use a full line between $a$ and $i$ whenever the
variable appearing in the clause is $x_i$ (i.e. $J^a_i=-1$), a dashed
line whenever the variable appearing in the clause is $\bar x_i$
(i.e. $J^a_i=1$).

Throughout this paper, the variable nodes indices are taken in
$i,j,k,...$, while the function nodes indices are taken in
$a,b,c,...$. For every variable node $i$, we denote by $V(i)$ the
set of function nodes $a$ to which it is connected by an edge, by
$n_i=|V(i)|$ the degree of the node, by $V_+(i)$ the subset
of $V(i)$ consisting of function nodes $a$ where the variable
appears un-negated --the edge $(a,i)$ is a full line--, and by
$V_-(i)$ the complementary subset of $V(i)$ consisting of function
nodes $a$ where the variable appears negated --the edge $(a,i)$ is a
dashed line--. $V(i)\setminus b$ denotes the set V(i) without a node $b$.
Similarly, for each function node $a$, we denote by
$V(a)=V_+(a)\cup V_-(a) $ the set of neighboring variable nodes,
decomposed according to the type of edge connecting $a$ and $i$,
and by $n_a$ its degree. Given a function node $a$ and a variable node $j$,
connected by an edge, it is also convenient to define the two sets:
$V^s_a(j)$ and $V^u_a(j)$, where the indices $s$ and $u$
respectively refer to the neighbors which tend to make variable
$j$ respectively satisfy or unsatisfy the clause $a$, defined as (see Fig.~\ref{fig_iter}):
\begin{eqnarray}
{\mbox{if}} \; && J_j^a=1 \; : \; \ V^u_a(j)= V_+(j) \ \ ; \ \
V^s_a(j)= V_-(j)\setminus a\\ {\mbox {if}}\; && J_j^a=-1: \; \
V^u_a(j)= V_-(j)\ \ ; \ \ V^s_a(j)= V_+(j)\setminus a
\end{eqnarray}

\section{Survey Propagation}
\label{wptosp}
The survey propagation algorithm (SP) is a message passing algorithm in which the
messages passed between function nodes and variable nodes have a probabilistic interpretation. Every message is the probability distribution of some warnings. In this section we provide a short self-contained introduction to SP, starting with the definition of warnings.
 We refer to refs. \cite{MPZ,MZ,BMZ} for  the original derivation of SP from 
statistical physics tools.

\subsection{Warnings}
The basic elementary message passed from one function node $a$ to a
variable $i$ (connected by an edge) is a Boolean number $u_{a \to i}
\in\{ 0,1\}$ called a `warning'.

 Given a function node $a$ and
one of its variable nodes $i$, the warning $u_{a \to i}$ is
determined from the warnings $u_{b \to j}$ arriving on all the
variables $j \in V(a)\setminus i$ according to:
\begin{equation}
u_{a \to i}= \prod_{j \in V(a)\setminus i} \theta\left(
 - J^a_j \left(\sum_{b \in V(j)\setminus a} J^b_j u_{b \to j} \right)
\right) \ ,
\label{udef}
\end{equation}
where $\theta (x) =0$ if $x \leq 0$ and $\theta (x) =1$ if $x > 0$.

A warning $u_{a \to i}=1$ can be interpreted as a
message sent from function node $a$, telling the variable $i$ that it
should adopt the correct value in order to satisfy clause $a$. This is
decided by $a$ according to the warnings which it received from all
the other variables $j$ to which it is connected: if $ \left(\sum_{b
\in V(j)\setminus a} J^b_j u_{b \to j} \right) J^a_j<0$, this means
that the tendency for site $j$ (in the absence of $a$) would be to
take a value which does not satisfy clause $a$. If all neighbors $j
\in V(a)\setminus i$ are in this situation, then $a$ sends a warning
to $i$.

The simplest message passing  algorithm one can think of is just the propagation of warnings,
in which the rule (\ref{udef}) is used as an   update rule which  is implemented sequentially.
The interest in WP largely comes from the fact that it gives the
exact solution for tree-problems.  In more complicated problems where the factor graph is not a tree, it turns out that 
WP is unable to converge whenever the density of constraints is so large that the solution space is clustered.
This has prompted the development of the more elaborate SP message passing procedure.

\subsection{The algorithm}

A  message of SP, called a  survey, passed from one function node $a$ to a
variable $i$ (connected by an edge) is a real
number $\eta_{a \to i} \in [0,1]$. 
The interpretation 
of the survey $\eta_{a \to i}$ is a probability among all clusters of satisfying assignments that a warning 
is sent from $a$ to $i$.

The SP algorithm
uses a random sequential update (see Alg.~\ref{SPalgo}),
which calls the local update rule SP-UPDATE (Alg.~\ref{SPsub}):

\begin{algorithm}
\caption{ : SP}
\label{SPalgo}
\INPUT{The factor graph of a Boolean formula in conjunctive normal
form; a maximal number of iterations $t_{max}$; a requested precision
$\epsilon$}
\OUTPUT{UN-CONVERGED if SP has not converged after $t_{max}$
sweeps. If it has converged: the set of all 
messages $\eta_{a \to i}^*$}
\begin{algorithmic}[1]
\STATE At time $t=0$: 
\FOR{every edge $a \to i$ of the factor graph}
\STATE randomly initialize the messages $\eta_{a \to i}(t=0) \in [ 0,1]$
\ENDFOR
\FOR{$t= 1$ to $t=t_{max}$}
\STATE sweep the set of edges in a random order, and update
 sequentially the warnings on all the edges of the graph,
 generating the values $\eta_{a \to i}(t)$, using subroutine
 BP-UPDATE
\IF{$|\eta_{a \to i}(t)-\eta_{a \to i}(t-1)|<\epsilon $
on all the edges}
\STATE the iteration has converged and generated $\eta_{a \to
i}^*=\eta_{a \to i}(t)$
\GOTO 12
\ENDIF
\ENDFOR
\IF{$t=t_{max}$}
\RETURN UN-CONVERGED
\ELSE[$t<t_{max}$] 
\RETURN the set of fixed point warnings $\eta_{a \to i}^*=\eta_{a \to i}(t)$
\ENDIF
\end{algorithmic}
\end{algorithm}

\begin{algorithm}
\caption{ : subroutine SP-UPDATE$(\eta_{a \to i})$}
\label{SPsub}
\INPUT{Set of all messages arriving onto each variable node $j \in
V(a) \setminus i$}
\OUTPUT{New value for the message $\eta_{a \to i}$}
\begin{algorithmic}[1]
\FOR{every $j \in V(a) \setminus i$}
\STATE compute the three
numbers:
\begin{eqnarray}
\nonumber \Pi^u_{j\to a}&=& \left[1-\prod_{ b\in
 V^u_a(j)}\left(1-\eta_{b\to j}\right)\right] \prod_{ b\in V^s_a (j)
 }\left(1-\eta_{b\to j}\right) \\ \nonumber \Pi^s_{j\to a}&=&
 \left[1-\prod_{ b\in V^s_a(j)}\left(1-\eta_{b\to j}\right)\right]
 \prod_{ b\in V^u_a (j) }\left(1-\eta_{b\to j}\right)\\ \Pi^0_{j\to
 a}&=& \prod_{ b\in V(j)\setminus a }\left(1-\eta_{b\to j}\right)
\label{eta2}
\end{eqnarray}
 if a set like $ V^s_a(j)$ is empty, the corresponding product takes
value $1$ by definition.

\STATE Using these numbers , compute and return the new survey:
\begin{equation}
 \eta_{a \to i}=\prod_{j \in V(a)\setminus i} \left[ \frac{\Pi^u_{j\to
a}} {\Pi^u_{j\to a}+\Pi^s_{j\to a}+\Pi^0_{j\to a}} \right] \ .
\label{eta1}
\end{equation}
If $V(a) \setminus i$ is empty, then $\eta_{a \to i}=1$.
\ENDFOR
\end{algorithmic}
\end{algorithm}

 Whenever the SP algorithm converges to
a fixed-point set of messages $\eta_{a \to i}^*$, one can use it 
in a decimation procedure in order to find a satisfiable assignment,
if such an assignment exists. This procedure, called the survey inspired 
decimation (SID) is described  in Alg.~\ref{SIDalgo}.

\begin{algorithm}
\caption{ : SID}
\label{SIDalgo}
\INPUT{The factor graph of a Boolean formula in conjunctive normal
form.  A maximal number of iterations $t_{max}$ and a precision
$\epsilon$ used in SP}
\OUTPUT{One assignment which satisfies all clauses, or 'SP
UNCONVERGED', or 'probably UNSAT'}
\begin{algorithmic}[1]
\STATE Random initial condition for the surveys
\STATE run SP  
\IF{SP does not converge}
\RETURN 'SP UNCONVERGED' and stop (or restart, i.e.\sbf{goto} line 1).  
\ELSE[if SP converges]
\STATE keep the fixed-point surveys $\eta^*_{a \to i}$; \sbf{goto} line 9
\ENDIF
\STATE Decimate:
\IF{non-trivial surveys ($\{ \eta \neq 0\}$) are
found} 
\FOR{each variable node $i$}
\STATE evaluate the three 'biases' $\{W_i^{(+)}, W_i^{(-)}, W_i^{(0)}\}$ defined by:
\begin{eqnarray}
W_i^{(+)} &=& \frac{\hat\Pi_i^+}{\hat\Pi^+_{i}+\hat\Pi^-_{i}+\hat\Pi^0_{i}} \label{wdef1}\\
W_i^{(-)} &=& \frac{\hat\Pi_i^-}{\hat\Pi^+_{i}+\hat\Pi^-_{i}+\hat\Pi^0_{i}} \label{wdef2}\\
W_i^{(0)} &=& 1-W_i^{(+)}-W_i^{(-)}
\label{wdef3}
\end{eqnarray}
where $\hat \Pi_i^+,\hat\Pi_i^-,\hat\Pi_i^0$ are defined by
\begin{eqnarray}
 \hat \Pi^+_{i}&=&
\left[1-\prod_{ a\in V_+(i)}\left(1-\eta^*_{a\to i}\right)\right]
\prod_{ a\in V_- (i) }\left(1-\eta^*_{a\to i}\right) \nonumber \\
\hat \Pi^-_{i}&=&
\left[1-\prod_{ a\in V_-(i)}\left(1-\eta^*_{a\to i}\right)\right]
\prod_{ a\in V_+ (i) }\left(1-\eta^*_{a\to i}\right) \nonumber \\
\hat \Pi^0_{i}&=&
\prod_{ a\in V(i) }\left(1-\eta^*_{a\to i}\right)
\label{hatpidef}
\end{eqnarray}

 \STATE\underline{fix} the variable with the largest $\vert
W_i^{(+)}- W_i^{(-)}\vert$ to the value $x_i=1$ if $W_i^{(+)}>
W_i^{(-)}$, to the value $x_i=0$ if $W_i^{(+)}< W_i^{(-)}$  
\STATE Clean the
graph, which means: toremove the clauses satisfied by this fixing,
to reduce the clauses that involve the fixed variable with opposite
literal, and to update the number of unfixed variables.
\ENDFOR
\ELSE[if all surveys are trivial ($\{ \eta=0\}$) ]
\STATE\sbf{output} the simplified sub-formula and run on it a local search process
(e.g. walksat).
\ENDIF

\IF{the problem is solved completely by unit clause
propagation}
\RETURN ``SAT'' and \sbf{stop}  
\ELSIF{no contradiction is
found}
\STATE continue the decimation process on the smaller problem (\sbf{goto} line 2)  
\ELSE[if a contradiction is reached] 
\RETURN 'probably UNSAT'
\STATE \sbf{stop}
\ENDIF
\end{algorithmic}
\end{algorithm}

There exist several variants of this algorithm. In the code which is
available at \cite{webserie}, for performance reasons we update
simultaneously all $\eta$ belonging to the same clause. The clauses to
be updated are chosen in a random permutation order at each iteration
step. The algorithm can also be randomized by fixing, instead of the
most biased variables, one variable randomly chosen in the set of the
x percent variables with the largest bias.  This strategy allows to
use some restart in the case where the algorithm has not found a
solution. A fastest decimation can also be obtained by fixing in the
step $12$, instead of one variable, a fraction $f$ of the $N_t$
variables (the most polarized ones) which have not yet been fixed (going back to $1$ variable
when $f N_t<1$).

In Ref.~\cite{BMZ}, experimental results are reported which show that 
SID is able to solve efficiently huge SAT instances very close to the threshold.
More specifically, the data  discussed in Ref.~\cite{BMZ}  show that if SID fixes at
each step only one variable, it converges in $\Theta(N^2 \log
N)$ operations (the time taken by walksat \cite{Selman} to solve the simplified
sub-formula seems to grow more slowly).  When  a fraction of
variables is fixed at each time step, we get a further reduction of the cost to $O(N
(\log N)^2)$ (the second $\log$ comes from sorting the biases).

In the  following sections we show how the SP procedures can be made fully
distributed and hence  amenable to a parallel implementation.

\section{Distributed SP  I: Diffusion algorithm}

\subsection{Simulating SP in parallel}

The SP algoriths described in the preceeding section 
 is by constructions distributed algorithms, since updates of nodes
are performed using message passing procedures from nearest neighbors. 
However when one uses the surveys in order to find the SAT assignments, in the SID procedure, the decimation
process breaks this local information exchange design: it requires
a global information, namely the maximally polarization field, 
used to decide which node has to be frozen at first. It is the purpose 
of this section to define a procedure able to diffuse such a global
information, by using the message passing procedure.
 
In the present  distributed implementation of SP (see Alg. \ref{DDalgo}), each node
is responsible for its own information, any sort of centralized
information is prohibited. 
The information stored at a given function node $a$ is 
the set of surveys $\{\eta_{a\to i}, i\in V(a)\}$. 
In turn, a given variable node $i$, keeps an 
information which is the set of cavity-fields 
$\{\Pi_{i\to a}, a\in V(i)\}$.
The scheme amounts to implement a message passing procedure
such that function nodes send their $\{\eta\}$ values to neigbouring variables 
nodes, and variable nodes send their fields
values in return. Each time a node receives a message from
its neighbours, it has to update its own information, accordingly
to the survey propagation equations.
The broadcasting  of information for each individual 
node is supposed to occur at random. 
Our simulation  proceeds by random updating of each variable or function
node. 
The update time-scale,
that is the rate at which nodes in a distributed implementation are able 
to collect new information and perform their own update will be denoted 
by $\tau$.  
Typically, as we shall see the full decimation process is of the
order of $10^4\tau$, which means that in 
average each node has to update $10^4$ time before the
process is completed. 

\begin{algorithm}
\caption{ : diffusive decimation algorithm}
\label{DDalgo}
\INPUT{The  factor graph of a Boolean  formula in conjunctive normal form;
a maximal computational time $t_{max}$, a requested precision $\epsilon$, a number $n_{su}$
of succesive  update with maximum polarization 
required to freeze a variable, the damping factor $\delta$ for information decay,
a random set of cavity biases $\eta_{a\to i}$.}
\OUTPUT{UN-CONVERGED if the algorithm has not converged after
$t_{max}$. If it has converged: one assignement which satisfy a fraction of the clauses, and a remaining 
reduced factor graph corresponding to a paramgnetic state which can be solved calling \textsc{walksat}}
\begin{algorithmic}[1]
\STATE At time $t=0$: 
\STATE  Initialize the counter $n_{su}(i)=0,i=1\ldots N$ 
\STATE parallel update of each variable node using VARIABLE-NODE-UPDATE.
\FOR{$t<t_{max}$ AND $H_{max}>\epsilon$}
\STATE parallel asynchronized 
update of each variable node
using VARIABLE-NODE-UPDATE and each function node using 
FUNCTION-NODE-UPDATE, at a mean rate of time $\tau$ per node.
\ENDFOR
\IF{$H_{max}<\epsilon$}
\RETURN the set of assignments of the frozen variables and the remaining reduced factor graph of unfrozen
variables. 
\ELSE
\RETURN UN-CONVERGED
\ENDIF
\end{algorithmic}
\end{algorithm}

\begin{algorithm}
\caption{ : subroutine VARIABLE-NODE-UPDATE}
\label{VNUalgo}
\INPUT{Set of all cavity bias surveys $\{\eta_{a\to i}, a\in V(i)\}$ arriving at a given
variable node $i$, information $H_{max}(a)$ on the maximum polarization comming from each neighbor $a\in V(i)$}
\OUTPUT{New value for the set of probabilites $\{\Pi^\pm_{i\to a},\Pi^0_{i\to a},a\in V(i)\}$, 
new information  $H_{max}(i)$
on the maximum polarization as estimated by $i$}
\begin{algorithmic}[1]
\FOR{every $a \in V(i)$} 
\STATE compute the values of $\Pi^\pm_{i\to a},
\Pi^0_{i\to a}$ using eq. (\ref{eta2}).
\ENDFOR
\STATE evaluate the two 'biases' $\{W_i^{(+)}, W_i^{(-)}\}$ using eq. (\ref{wdef1},\ref{wdef2}):
\IF{$H_{max}(a) < \vert W_i^{(+)}- W_i^{(-)}\vert +\epsilon $}
\STATE $H_{max}(i) = \vert W_i^{(+)}- W_i^{(-)}\vert$
\STATE $n_{su}(i) = n_{su}(i)+1$
\IF {$n_{su}(i)> n_{su}$}
\STATE freeze variable $i$ to the value $x_i=1$ if $W_i^{(+)}>
W_i^{(-)}$, to the value $x_i=0$ if $W_i^{(+)}< W_i^{(-)}$
\ENDIF
\ELSE
\STATE $H_{max}(i) = \text{Max}(H_{max}(i),H_{max}(a)) * (1-\delta)$
\STATE $n_{su}(i) = 0$
\ENDIF
\end{algorithmic}
\end{algorithm}

\begin{algorithm}
\caption{ : subroutine FUNCTION-NODE-UPDATE}
\label{FNUalgo}
\INPUT{Set of all probabilites $\{\Pi^\pm_{i\to a},\Pi^0_{i\to a},i\in V(a)\}$ arriving at a given
function node $a$, information $H_{max}(i)$ on the maximum polarization comming from each neighbor $i\in V(a)$}
\OUTPUT{New value for the set of cavity bias survey $\{\eta_{a \to i}, i\in V(a)\}$, new information  $H_{max}(a)$
on the maximum polarization as estimated by $a$}
\begin{algorithmic}[1]
\FOR{every $i \in V(a)$} 
\STATE Compute  $\eta_{a \to i}$ using eq. (\ref{eta1}).
\STATE $H_{max}(a) = \text{Max}(H_{max}(i),H_{max}(a)) * (1-\delta)$
\ENDFOR
\end{algorithmic}
\end{algorithm}

\subsection{Propagation of information}
Messages can contain additional information to the 
one needed to run \textsc{sp}.  In the present case,
the decimation procedure diffuses the 
information concerning the highest polarization fields $H_{max}=\text{Max}(|W^+ - W^-|)$
in the system, in order that the variable node, being aware to be the most polarized one, can decide 
to freeze its assignment to the one  given by the orientation 
of the polarization field. 

A static information can freely 
propagate  by simply diffusing in the network, and 
any component of the network should be informed by 
a time which has to be proportional to $\log(N)$, $N$ being the size (number of variables nodes) of the 
system. The difficulty is that the information to be broadcasted may still vary with time
before the system has reached equilibrium. The 
difficulty in propagating a signal saying that the system has
converged, might be even more severe.
Indeed this amounts to propagate the maximum displacement from
the last update of each individual node, which is by essence
a transient information. 

To take advantage of the  basic message-passing procedure, 
in the case in which the information to be diffused can vary 
with time, we have to add a mechanism able to suppress 
obsolete information.
The way we propose to  
do this is to incorporate some damping factor in the 
broadcasted values. 

Let us call
$\delta\ll 1$ this damping factor. The procedure goes as follows (see Alg. \ref{VNUalgo}) 
Each node ($a$ for a function node or $i$ for a variable node) stores its  own estimate of  
the max polarization field $H_{max}$ in a local variable $H_{max}(i)$ or $H_{max}(a)$.
Consider first a variable node $i$. When this updates, in addition to the
set $\{\eta_{b\to i}\}$ from its neighbour function nodes $b$, it collects
also the information $H_{max}(b)$ about the maximum, coming from these
nodes, compare it to its own estimation $H_{max}(i)$, and also 
to its own field value $W_i=|W_i^+-W_i^-|$. In case the new $H_{max}(i)$ estimation
indicates that $i$ is not the most polarized variable node ($H_{max}(i)>W_i$), 
$H_{max}(i)$ is multiplied by a factor $(1-\delta)$. If instead,
it appears that $i$ might be the most polarized node ($H_{max}(i) = W_i$), 
then $H_{max}(i)$ is kept as it stands, with no damping correction. For a
function node, when it updates, we apply this damping factor anyway.
The virtue of this damping effect, is to eliminate
any false information. Suppose indeed, that at a given time,
a strongly polarized variable node diffuses in the network a very high 
value of $H_{max}$; if after some time  this value is not confirmed (because
this node converges to a less polarized state), which means that all local fields $W_i$ will be 
smaller than $H_{max}$, and no unit in the network can pretend to be the most polarized one. Then, mechanically  
the estimation on the maximum will decay at a rate $\delta$, 
until when a variable node 
presents a polarization field higher than the decaying false information.\\
The counterpart is that the damping takes effect spatially on the network.
So typically, since the radius of the network scales like $\log(N)$,
the distribution of the estimated maximum given by the set 
$\{H_{max}(i),H_{max}(a)\}$ taken over the network, will be enlarged by some
factor $\delta\log(N)$.\\
As a result when playing with the  parameter $\delta$, two contradictory
effects are at work:
\begin{itemize}
\item either the information diffuses rapidly with a poor precision 
(large $\delta$),
\item either it takes a long time, before a precise information is diffused (small $\delta)$.
\end{itemize}
When it comes to the decimation procedure, this can be 
actually turned into an advantage, since it provides us directly
with a parameter able to fix, in an heuristic manner, the rate 
at which the decimation is performed. Each variable node
is equipped with a counting variable $n_{su}(i)$ initially set to zero and counting successive
updates with the following two conditions:
\begin{itemize}
\item the variation since last updtate of the $\Pi_i$'s
is less than $\epsilon$ (convergence criteria, typically $\epsilon\simeq 10^{-3}$),
\item $i$ may consider itself to be the most polarized variable ($H_{max}(i) = \vert W_i^+-W_i^-\vert$).
\end{itemize}
When one of these conditions is not satisfied, $n_{su}(i)$ is set to zero. If $n_{su}(i)$
is  greater than a parameter $n_{su}(i)$, the variable
is allowed to freeze in the direction given by $W_i^+-W_i^-$.  

\begin{figure} 
\begin{center}\includegraphics[scale=0.5]{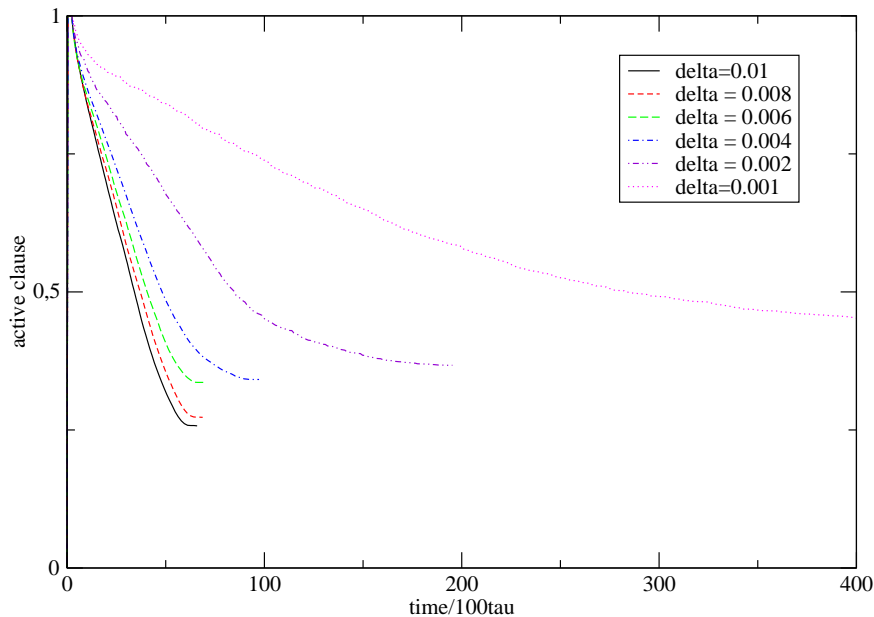}
\caption{Number of active clauses as a function of time for
different values of $\delta$ for a 3-\textsc{sat} problem
of size $N=10^5$ at $\alpha=4.2$ and $n_{su}=20$.}
\label{cyr2}
\end{center}
\end{figure}

\subsection{Experimental results}
We have simulated  the behavior of the 
algorithm using a standard computer, which allows to
monitor how much progress in computing time would be brought by its
fully parallel  implementation, and to optimize the choices of parameters.

\begin{figure}
\begin{center}
\includegraphics[scale=0.4]{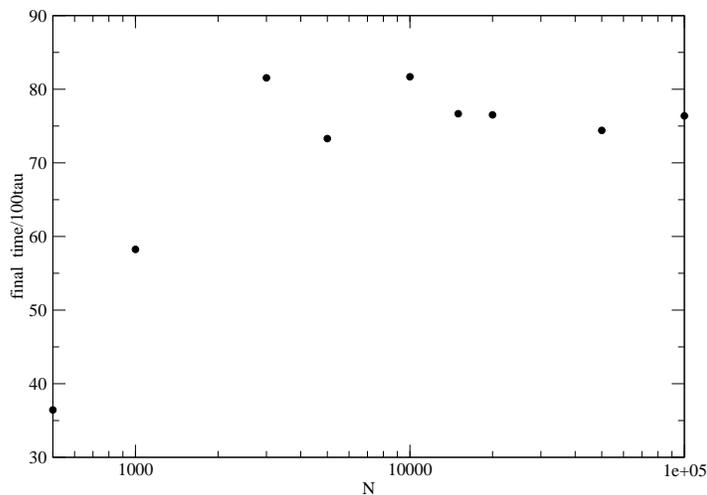}
\caption{Decimation time up to the paramagnetic state as a function
of the size of the system, for a 3-\textsc{sat} problem
at $\alpha=4.2$, $\delta=0.005$ and $n_{su}=20$.}
\label{cyr1}
\end{center}
\end{figure}

\begin{figure}
\begin{center}
\includegraphics[scale=0.4]{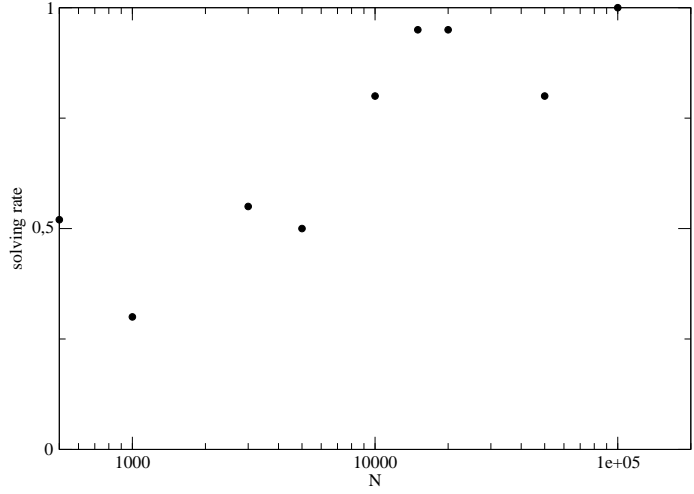}
\caption{Rate of succesfull decimation process as a function of
N for a 3-\textsc{sat} problem
at $\alpha=4.2$, $\delta=0.005$ and $n_{su}=20$.}
\label{cyr3}
\end{center}
\end{figure}

\begin{figure} 
\vspace{0.5cm}
\begin{center}
\includegraphics[scale=0.4]{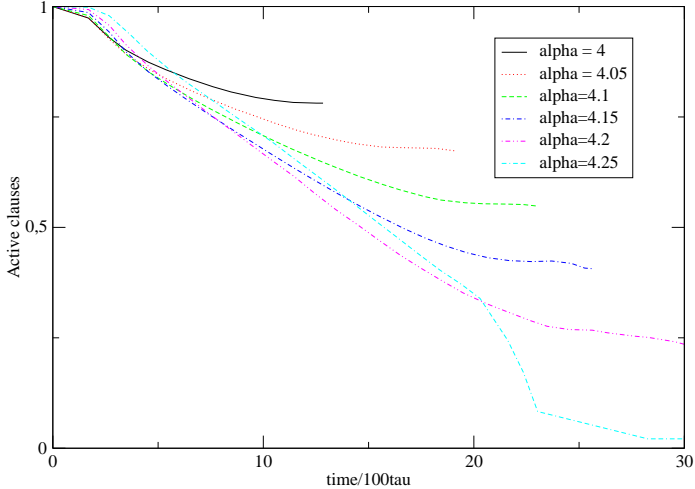}
\caption{Number of active clauses as a function of time 
during the decimation process for
different values of $\alpha$, for a 3-\textsc{sat} problem
of size $N=10^5$ at $\delta=0.01$ and $n_{su}=10$.}
\label{cyr4}
\end{center}
\end{figure}

 We use the random \textsc{3-sat} formulas as a benchmark problem.
Let  N denote the number of variables nodes and $\alpha=M/N$ the ratio of 
the number of function nodes $M$ over the
number of variable nodes. The range of interest is $\alpha\in[4,4.25]$.
Simulations are performed with  samples of $10^5$ function nodes.
Figures (~\ref{cyr2},~\ref{cyr1},~\ref{cyr3},~\ref{cyr4}) present some average over numerous decimation runs,
with various parameters. The runs stop when a paramagnetic state
is reached, or when the entropy becomes negative.
In order to assess the performance of the algorithm, we   measure 
the simulated time (the time it would take if the computation was
really distributed) to reach a paramagnetic state, which 
can be handled afterwards by \textsc{walksat} in order to find a solution.

The observables which we consider are
\begin{itemize}
\item $N_a$ the number of active (unfrozen)  variables nodes.
\item $M_a$ the number of active   function nodes.
\item $\alpha_2$ the ratio $M_{a2}/N_a$ where $M_{a2}$ is the number
of active two-clauses.
\item $\alpha_3$ the ratio $M_{a3}/N_a$ where $M_{a3}$ is the number
of active three-clauses. Since we start from a $3-$SAT problem we have of 
course $M_a=M_{a2}+M_{a3}$.
\item the entropy of the system, computed from the expression given in \cite{BMZ}
\end{itemize}

\subsubsection{Orders of magnitude for the tunning parameters}
The algorithm depends on two  tunable parameters
\begin{itemize}
\item the damping factor  $\delta$ on the diffusion of information. We find
that $\delta\in[0.001,0.01]$ is roughly the range of effectiveness of this parameter for $N=10^5$.
This actually gives essentially the decimation
rate between two convergences of \textsc{sp}.
For $\delta\geq 0.01$, too many variables ($\simeq 10^3$) are frozen at the same time, and 
the algorithm ceases to find a paramagnetic state.
Instead, when $\delta<0.001$, variables are decimated one by one. The time-dependence of the decimation process 
with this parameter is illustrated in fig. \ref{cyr2}.  
\item $n_{su}$ the number of successive successful updates 
needed for a variable, to get  frozen. It depends  on the precision $\epsilon$ which is required.
For $\epsilon=10^{-3}$ it is found that the range of validity of this parameter is between   
$10$ and  $50$. Above $50$, \textsc{sp} converges
before the next decimation cascade. Below $20$ the decimation process
is essentially continuous. Below $10$ the algorithm ceases to converge to a paramagnetic
state. 
\end{itemize} 

A reasonable choice for the selection of parameters leading to a 
fast but still successful  decimation procedure is around  $\delta=0.01$
and $n_{su} = 10$.

\subsubsection{Dependence with the size of the system}
Once the typical value for the tunable parameter are determined, it is possible to
study the behavior of the algorithm with the sixe of the problem, $N$. Fig. \ref{cyr1},
shows that the dependence in $N$ is of the order of $\log N$, 
which is the best performance one could have with such an algorithm.
Another expected result is that rate of successful decimation should increase 
when  $N$ increases, since finite size effect are known to alter efficiency 
of algorithms for solving \textsc{3-sat} formulas. This is what is effectively 
observed in fig. \ref{cyr3}

\subsubsection{Dependence with $\alpha$}
\begin{figure} 
\begin{center}\includegraphics[scale=0.4]{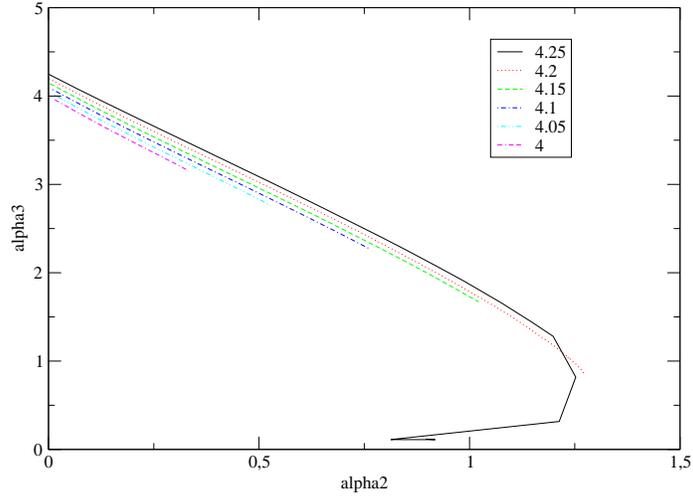}
\caption{Trajectories in the phase space $(\alpha_2,\alpha_3)$
during the decimation process for various values of
$\alpha$
for a 3-\textsc{sat} problem
of size $N=10^5$ at $\delta=0.01$ and $n_{su}=10$.
($5$ samples for each value of $\alpha$, all decimations processes where 
succesfull up to $\alpha=4.2$, and all failed for $\alpha=4.25$) The
slope of the linear part of the trajectories is $\simeq -2.4$.
}
\label{cyr5}
\end{center}
\end{figure}

Depending on $\alpha$ we observe that when we approach the transition point $\alpha_c\simeq 4.26$,
the final decimation rate increases, (see fig. \ref{cyr4}) which is consistent with the hypothesis
that the size of the backbone of  solutions clusters increases when we approach the critical value of $\alpha$.
This can be also visualized when looking at the phase space trajectory of the decimation process (see
fig. \ref{cyr5}. During the process 2-clauses are generated, which leads to represent the trajectory
in the plane  $(\alpha_2,\alpha_3)$. The trajectory is shorter when we go away from 
$\alpha_c$. 

\section{Distributed SP II: 'Reinforcement algorithm'}

\subsection{SP with external forcing field}

The SP procedure has recently been generalized to allow the retrieval of a large number of different solutions thanks to the introduction of \textit{probability preconditionings}~\cite{BBCZ,BBCZ2}. 
The original factor graph is modified applying onto each variable $i$ a constant external forcing survey in direction $\xi_i\in\{-1,1\}$ and intensity $\eta_i^{ext} = \pi$. An \textit{a priori} probability $\pi$ of assuming the value $\pm1$ is being thus assigned to the variable $i$. The components of the forcing modify the usual SP equation (\ref{eta2}) in the following way~:   
\begin{eqnarray}\label{PIeq}
\nonumber
 \Pi^\pm_{j\to a}&=&
\left[1-\left(1-\pi\delta_{\xi_j,\pm}\right)\prod_{ b\in V_\pm(j)\setminus a}\left(1-\eta_{b\to j}\right)\right]\cdot\nonumber\\
&&\cdot\left(1- \pi\delta_{\xi_j,\mp}\right)\prod_{ b\in V_\mp (j) }\left(1-\eta_{b\to j}\right) \\
 \Pi^0_{j\to a}&=&\left(1-\pi\delta_{\xi_j,+}\right)\left(1-\pi\delta_{\xi_j,-}\right)\prod_{ b\in V(j)\setminus a }\left(1-\eta_{b\to j}\right)\nonumber
\end{eqnarray}

These relations, together with (\ref{eta1}), form the SP equations with external messages (SP-ext). The quantities $\hat \Pi_i^+,\hat\Pi_i^-,\hat\Pi_i^0$ used to define the local biases on each variable $i$ (see Eq.~\ref{hatpidef}) are modified in a similar manner and expressed as a function of the new probabilities $\eta^*_{b\to j}$ at the fixed point~:
\begin{eqnarray}
\nonumber
 \hat\Pi^\pm_{j\to a}&=&
\left[1-\left(1-\pi\delta_{\xi_j,\pm}\right)\prod_{ b\in V_\pm(j)}\left(1-\eta^*_{b\to j}\right)\right]\cdot\nonumber\\
&&\cdot\left(1- \pi\delta_{\xi_j,\mp}\right)\prod_{ b\in V_\mp (j) }\left(1-\eta^*_{b\to j}\right) \\
 \hat\Pi^0_{j\to a}&=&\left(1-\pi\delta_{\xi_j,+}\right)\left(1-\pi\delta_{\xi_j,-}\right)\prod_{ b\in V(j) }\left(1-\eta^*_{b\to j}\right)\nonumber
\label{hatpiext}
\end{eqnarray}

The external messages drive the value of the surveys $\eta^*_{a\to i}$ towards the clusters maximally aligned with the external vector field. $\pi=0$ corresponds to the case in which the external messages are absent ; in the case $\pi=1$, the system is fully forced in the direction $\vec{\xi}$. When $\pi$ is correctly chosen, this enables to address more specifically (compared to the SID algorithm, which permits to retrieve only one solution) clusters close to a given region of the $N$-dimensional space. Thus, a right choice of the \textit{a priori} probability $\pi$, and the use of alternative random directions $\vec{\xi}$ invariably permit to retrieve many solutions in many different clusters. 

The criterion used to decimate towards a solution is still the global criterion of the maximum of the magnetization over all variables. So, the decimation based on the SP-ext equations is not yet a distributed solver of the K-SAT formulas.

\subsection{Presentation of the method}

In order to make the decimation procedure fully distributed, we may adopt time-dependent external messages which are updated according to a reinforcement strategy. More precisely, good properties of convergence are obtained adopting the following scheme~: 
\begin{itemize}
\item in one time step (sweep), all the $\eta$s are parallely updated using the SP equations with external messages(\ref{eta1}, \ref{PIeq}), which in this scheme are time-dependent,
\item every second time step, one calculates all temporary local fields  $\{W_i^{(+)}, W_i^{(-)}, W_i^{(0)}\}$ using the equations (\ref{wdef1}, \ref{wdef2}, \ref{wdef3}, \ref{hatpiext}) ---note that these are also functions of the external messages---. The directions $\xi_i,\:\forall i\in\{1,\ldots,\:N\}$ of the temporary local messages, which are now time-dependent, are parallely updated to equal $sign(W_i^+-W_i^-)$, \textit{i.e.} to align itself with the local bias $W_i^+ - W_i^-$.
\end{itemize}

As we shall see, with an appropriate choice of the intensity $\pi$ of the forcing,  most variables are completely polarized at the end of a single convergence. A solution of the K-SAT formula is finally found by fixing each variable in the direction of its local bias. 
The reinforcement algorithm (RA) defined above is purely local and hence gives an efficient distributed solver for random K-SAT formulas in the hard-SAT phase.

The algorithm is precisely described in Alg.~(\ref{SPIIalgo}). Its interpretation and  a preliminary  study of some of its properties are discussed in
the  subsequent sections.

\begin{algorithm}
\caption{ : reinforcement algorithm}
\label{SPIIalgo}
\INPUT{The  factor graph of a Boolean  formula in conjunctive normal form;
a maximal number of iterations $t_{max}$, a requested precision $\epsilon$ and the
probability $\pi$ that the external message sends a warning}
\OUTPUT{UN-CONVERGED if the algorithm has not converged after
$t_{max}$ sweeps. If it has converged: one assigment which satisfies all clauses}
\begin{algorithmic}[1]
\STATE At time $t=0$: 
\FOR{every edge $a \to i$ of the factor graph}
\STATE randomly initialize the cavity biases $\eta_{a \to i}(t=0) \in [ 0,1]$. The additional surveys are initially set to $Q_i(\vec{u_i},0)$
\ENDFOR
\FOR{$t= 1$ to $t=t_{max}$}
\STATE update parallelely the cavity bias surveys on all the edges
of the graph, generating the values $\eta_{a \to i}(t)$, using subroutine
CBS-UPDATE.
\STATE half of the times, update parallelely the reference direction $\xi_i\in\{-1,1\}$ for all variables $x_i$, using subroutine FORC-UPDATE.
\IF{$|\eta_{a \to i}(t)-\eta_{a \to i}(t-1)|<\epsilon$ on all the edges}
\STATE the iteration
has converged and generated $\eta_{a \to i}^*=\eta_{a \to i}(t)$
\GOTO 13
\ENDIF
\ENDFOR
\IF{$t=t_{max}$}
\RETURN UN-CONVERGED
\ELSE[If $t<t_{max}$]
\RETURN the satisfying assignment which is obtained by fixing the boolean variable $x_i$ parallel to its local field~: $x_i = sign\{W_i^{(+)}-W_i^{(-)}\}$, and by performing the simplifications over the neighboring variables
\ENDIF
\end{algorithmic}
\end{algorithm}

\begin{algorithm}
\caption{ : subroutine CBS-UPDATE$(\eta_{a \to i})$}
\label{CBSalgo}
\INPUT{Set of all cavity bias surveys arriving onto each
variable node $j \in V(a)\setminus i$}
\OUTPUT{New value for the cavity bias survey $\eta_{a \to i}$}
\begin{algorithmic}[1]
\FOR{every $j \in V(a)\setminus i$} 
\STATE compute the values of $\Pi^\pm_{j\to a},
\Pi^\mp_{j\to a},\Pi^0_{j\to a}$ using eq. (\ref{PIeq}).
\STATE Compute  $\eta_{a \to i}$ using eq. (\ref{eta1}).
\ENDFOR
\end{algorithmic}
\end{algorithm}

\begin{algorithm}
\caption{ : subroutine FORC-UPDATE$(\xi_i)$}
\label{FORCsub}
\INPUT{Set of all cavity bias surveys arriving onto the
variable node $i$, including the additional survey $Q_i(\vec{e}_{\xi_i};\pi)$}
\OUTPUT{New value for the direction $\xi_i$ of the additional survey}
\begin{algorithmic}[1]
\STATE Compute the local fields $W_i^{(+)}, W_i^{(-)}, W_i^{(0)}$ using eq. (\ref{wdef1}, \ref{wdef2}, \ref{wdef3}, \ref{hatpiext})
\STATE Compute $\xi_i = sign(W_i^{(+)} - W_i^{(-)})$
\end{algorithmic}
\end{algorithm}

To understand how the reinforcement strategy is working, one has to look back at the interpretation of the SP equations with an external forcing survey~: imposing an external forcing survey of direction $\xi_i\in\{-1,1\}$ on the variable $i$ tries to drive the solution of the cavity biases distributions towards the clusters with a local orientation matching the externally imposed direction $\xi_i$.

Then, choosing $\vec{\xi}$ parallel to the vector of the temporary local fields is a strategy which typically guarantees that, if the intensity $\pi$ of the forcing is small enough, the overlap between the forcing vector field $\vec{\xi}$ and the closest solution is continuously increasing, in a stronger way than in the case of the usual SP equations.

When $\pi$ is correctly chosen, some satisfying assignments tend to become stable fixed points of the RA dynamics. The optimal value of $\pi$ is a trade-off between the need for convergence of the SP equations with external forcing survey, the strength of the stabilization induced on the satisfying assignments by the update rule and the amplitude of the perturbation brought by the external forcing. There is \textit{a priori} no guarantee that there exists a particular value of the intensity $\pi$ for which the algorithm, starting from random initial conditions, gets trapped by a single cluster, and eventually converges to a single solution inside this cluster, without any further decimation. But, our experimental results show that, for $N$ big enough and for almost all values of the parameter $\alpha$ corresponding to the SAT phase, there exists an intensity $\pi$ for which the overlap between the set of local fields and the closest satisfying assignment increases until when it eventually reaches $1$ at the end of a unique convergence.

\subsection{Results}

\subsubsection{Determination of the optimal intensity of the forcing $\pi$}

\begin{figure}
\begin{centering}
\includegraphics[scale=0.5,angle=270]{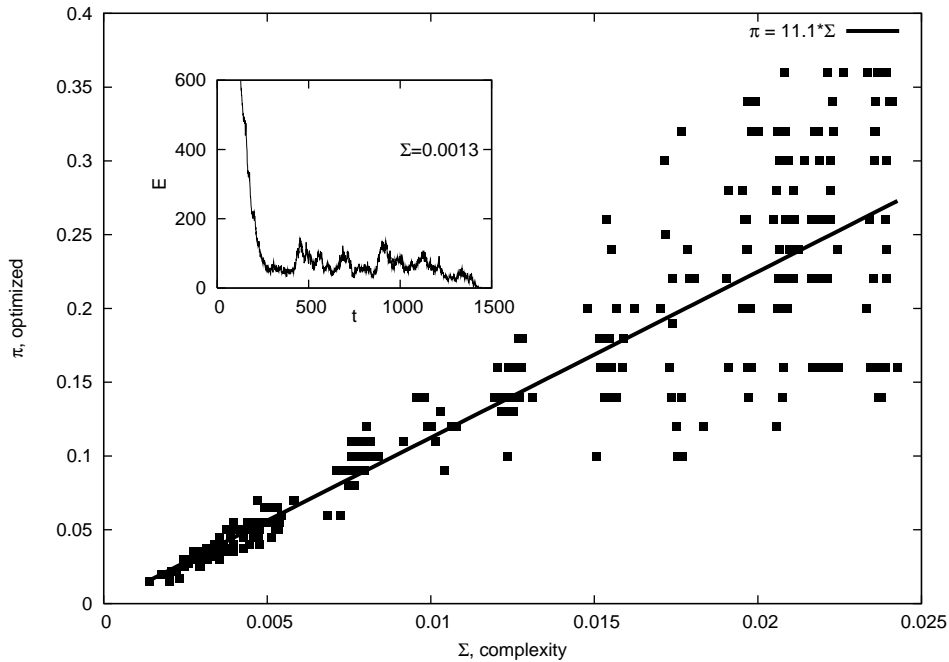}
\caption{Scaling of the optimized $\pi$ against the complexity $\Sigma$. The optimized $\pi$, which is the intensity of the forcing permitting to find a satisfying assignment in the minimum number of steps, is recalculated for each formula~: it depends in average linearly on the complexity~: $\pi=11.1\times\Sigma$ ($N=10^5$,  $K=3$, $4.0 < \alpha < 4.24$). Each point corresponds to a different formula~; note that the relationship $\pi_{opt} = f(\Sigma)$ becomes slightly sublinear for $\Sigma < 0.003$. The subplot shows the evolution of the energy of the dump solution (in which each variable is fixed parallel to the temporary local field)
that would be obtained if the convergence was stopped at time $t$ ($\alpha=4.24$). }\label{jo1}
\end{centering}
\end{figure}

The efficiency of the method is tested on random 3-SAT formulas. One first determines the optimal intensity $\pi_{opt}$, which is the intensity of the forcing which permits to find a satisfying assignment in the minimum number of steps. When the density of ground states clusters increases (for lower values of $\alpha$), $\pi_{opt}$ should be set at a higher value to be able to bias the convergence towards a unique cluster~; however, a too high value of $\pi$ may let the algorithm converge too rapidly in different parts of the graph towards assignments belonging to different clusters. The algorithm would then loop without convergence among these clusters. For these two reasons, $\pi_{opt}$ should be positively correlated with the complexity $\Sigma=\frac{log(N_{Cl})}{N}$. 

One observes that the  reinforcement process  diverges if the parallel update of the forcings is performed at a too high frequency with respect to the frequency of the parallel updates of the $\eta$s~; the local loops in which the algorithm gets trapped are likely due to simultaneous convergence towards different satisfying assignments in different parts of the graph. A good strategy, systematically used for the experiments of the present paper, is to perform the parallel updates of the forcings once every second parallel update of the $\eta$s.   

In the hard-SAT phase, one determines $\pi_{opt}$ as a function of the calculated complexity~\cite{MZ}. As predicted, the experiments show that, for $\alpha_D<\alpha<4.245$, $\pi_{opt}$ is positively correlated, and in average scales linearly, with $\Sigma$ ($\pi_{opt}=11.1\times\Sigma$, c.f. Fig.~\ref{jo1}, each point representing a different formula), and that the probability of converging towards a satisfying assignment using this value tends to $1$ ($N=10^5$, $\epsilon=10^{-3}$, $t_{max}=1000$, c.f. Fig.~\ref{jo2}).

\subsubsection{Range of validity of the method}

\begin{figure}
\begin{centering}
\includegraphics[scale=0.5,angle=270]{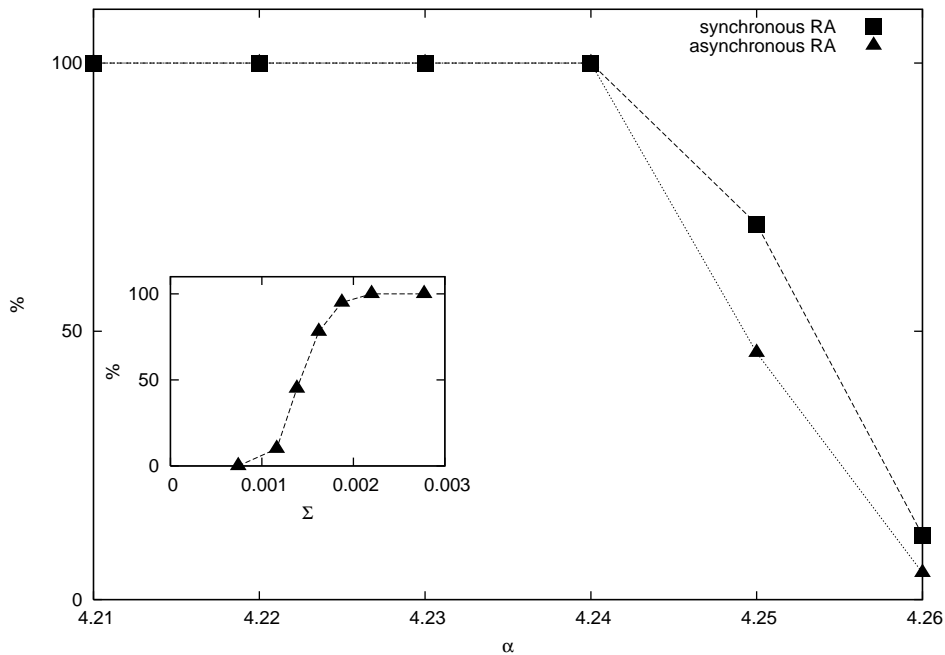}
\caption{Percentage of successful runs (convergence leading to a satisfying assignment) as a function of $\alpha$. The subplot shows the same percentage as a function of the calculated complexity $\Sigma$. The cut-off complexity, for which the RA is successful in half of the cases tested, is $\Sigma_{1/2} = 0.0013$ ($N=10^5$,$\epsilon=10^{-3}$,$t_{max}=1500)$. The range of validity of the reinforcement algorithm thus matches the one of the SID (see text).}\label{jo2}
\end{centering}
\end{figure}

In this section we compare RA with the so-called serialized survey propagation-inspired decimation (SID).

The figure \ref{jo2} plots the percentage of successful runs (\textit{i.e.} of convergence towards a satisfying assignment after a single run) of both the synchronous and the asynchronous RA as a function of $\alpha$. One calls synchronous RA the algorithm previously described (Alg.~\ref{SPIIalgo}) in which the update of all $\eta$s, as well as respectively the update of all forcings, are performed simultaneously. Conversely, in the asynchronous RA, the updates are performed sequentially in a random permutation order (note, that, in the latter case, forcings can be updated as frequently as the $\eta$s). For $\alpha=4.24$, both algorithms invariably permit to retrieve a solution. This is in accordance with the result presented in Ref.~\cite{BMZ}, in which the classical SID has been shown to always find a solution by fixing $0.125\%$ of variables after each convergence of the SP equations. 

The cut-off value of the complexity $\Sigma$, for which the decimation converges in half of the cases, is $\Sigma_{1/2}=0.0013$ (subplot of Fig.~\ref{jo2}). $\Sigma_{1/2}$ was constant for all sizes of the graph $N>2.10^4$. When the complexity decreases, the decimation progressively fails. Depending on the value of $\pi$, either it converges towards a set of local fields, which corresponds to a weighted average of the contributions of numerous clusters of solutions~; or, it loops among low energy configurations (usually, for $N=10^5$, $\alpha=4.25$, one finally finds assignments of energy $E=0-50$).

Moreover, asymptotically, the numerical critical $\alpha$ above which the SID, without backtracking, should not permit to find a satisfying assignment has been estimated to be $\alpha_A=4.252$~\cite{P1}, substantially smaller than the theoretical critical value $\alpha_C=4.2667$  (for the improvement brought by the backtracking procedure, see \cite{P2,BMZ}). 

One may study the behavior of the RA at $\alpha_A$ for $N=10^6$. The calculated complexity of the formulas with $N=10^6$ and $\alpha=4.252$ equals $\Sigma=0.00133\pm 0.00013$, in agreement with the cut-off complexity determined for formulas of size $N=10^5$ (Fig.~\ref{jo2}). For $N=10^6$ and $\alpha=4.252$, one then sets $\pi_{opt}=10.5\times \Sigma$ (in agreement with the curve of Fig.~\ref{jo1} for $N=10^5$). The ``reinforcement algorithm'' using these parameters nicely permits to converge towards a solution in $10$ out of the $15$ formulas tested. 

One concludes that the present algorithm is valid at least in the same range of parameters as the classical SID~\cite{BMZ}.  

\subsubsection{Which solutions are addressed?}
The satisfying assignments found by the present algorithm are highly dependent on the initial conditions, \textit{i.e.} on the initial random values imposed on the $\eta$s. For different random initial conditions, one converges towards solutions belonging to different clusters. 

To know where the found solutions are located, one chooses an instance from a random graph ensemble in which the number of clauses where any given variable appears negated or directed are kept strictly equal (\textit{bar-balanced formulas}). For such an instance ($N=30000$, $\alpha=3.33$), the clusters of solutions are expected (and have been observed) to be uniformly distributed over the whole N-dimensional space~\cite{BBCZ2}. The RA is launched in such an instance $800$ times starting from different random conditions. The histogram of the Hamming distances $d(S_1,S_2)=\frac{1}{2}(1-\frac{1}{N}\sum_ix_i^{(S_1)}x_i^{(S_2)})$ between any two of the 800 solutions peaks at $0.4995$ (with a standard deviation of $0.0118$).

There is no guarantee that the present algorithm is able to address all clusters of solutions, but the fact that the distance distribution among the found solutions of a bar-balanced formula peaks at $0.5$ with a standard deviation expected from a random distribution (\textit{i.e.} $\frac{2}{\sqrt{N}}=0.0115$), suggests that the addressable solutions could be homogeneously distributed over the space of the clusters of solutions.

\subsubsection{Logarithmic dependence of time on the size of the graph}

\begin{figure}
\begin{centering}
\includegraphics[scale=0.5,angle=270]{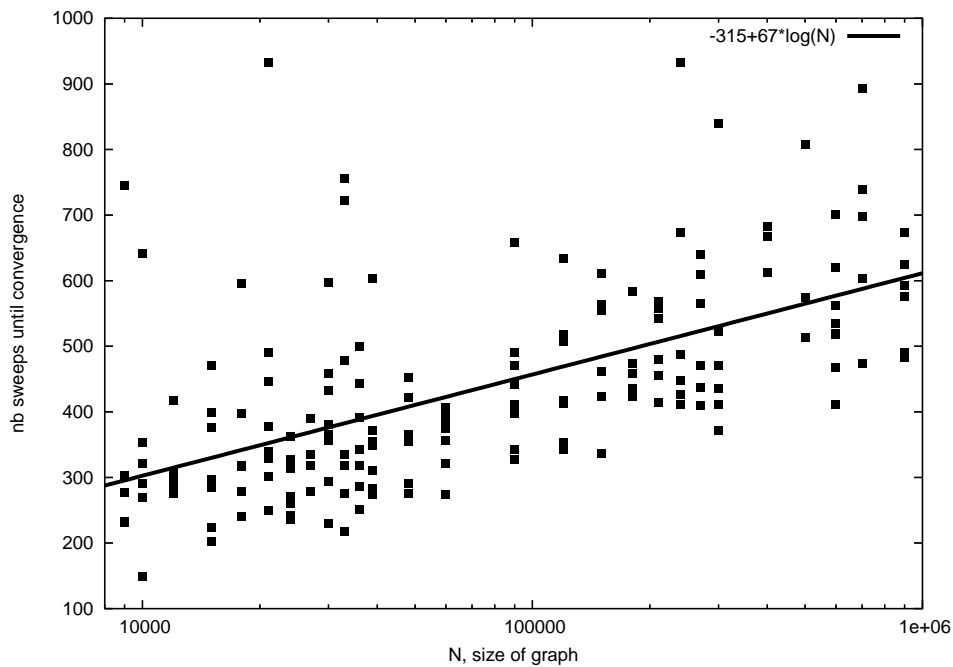}
\caption{The number of parallel sweeps needed to find a solution depends logarithmically on $N$ ($\alpha=4.22$, $\pi=0.04$). The examples in which the convergence fails (due to finite size effects, and which have not been observed for $N>30000$) are not represented.}\label{jo3}
\end{centering}
\end{figure}

One now analyses the evolution of the convergence time as a function of the size of the graph. For this, one determines, for a fixed value of $\alpha=4.22$, the simulated time that the convergence would take if implemented on a distributed device, as a function of the size $N$ of the graph (Fig.~\ref{jo3}). For $N$ ranging from $9000$ to $9.10^5$, the time of convergence is well approximated by a logarithmic fit [$t=-315+67*log(N)$]. The unit of time is one parallel update of all $\eta$s. Each point corresponds to a different random formula. Note that, above $N>3.10^4$, all trials were successful, and the RA invariably leads to a satisfying assignment. For smaller graphs, due to finite-size effects, the algorithm either fails to converge or converges to an assignment with a few contradictions (generally 1 to 4)  in few cases: at most $5\%$ of the experiments for $N=9000$ and none for $N>30000$.

As local messages need a logarithmic time to reach all parts of the graph, and as, in the hard-SAT phase, the assignment of a value to a given variable $i$ does not depend on purely local constraints but on the state of the whole graph, one expects that, by the use of purely local messages on a distributed device, the problem can not be solved faster than in a logarithmic time. 

\subsubsection{Performance of the  reinforcement algorithm on a serialized implementation}

Numerical experiments show that RA scales as $N\times\log(N)$ when implemented on a serialized computer. In the scheme of the classical SID, a slightly slower scaling, in $N\times(\log(N))^2$, can be achieved by fixing after each convergence a given percentage of the total amount of variable~\cite{BMZ}. 

Moreover, Ref.~\cite{BMZ} reports that, for $\alpha=4.24$ and $N=10^5$, by fixing at least $f=0.125\%$ of the $N$ variables at the end of each convergence, the SID reaches a paramagnetic state after an average of $7460$ sweeps, \textit{i.e.} $7460$ updates of all remaining $\eta$s. Conversely, the synchronous RA only needs $600$ parallel updates of the $\eta$s plus $300$ parallel updates of the forcings to converge towards a solution. Its asynchronous version requires $450-500$ complete updates of the $\eta$s and of the forcings to arrive at a solution. This means that, in a serialized computer, close to the critical point, the present algorithm performs faster than the classical SID by a factor of $5$. This can be checked by running the two algorithms on a 2.4 GHz PC~: the formula ($N=10^5$ and $\alpha=4.24$) is solved in average in $30$ minutes using the SID with $f=0.125\%$ and in $7$ minutes using the ``reinforcement algorithm'' (either asynchronous or synchronous). 

\begin{figure}
\begin{centering}
\includegraphics[scale=0.5,angle=270]{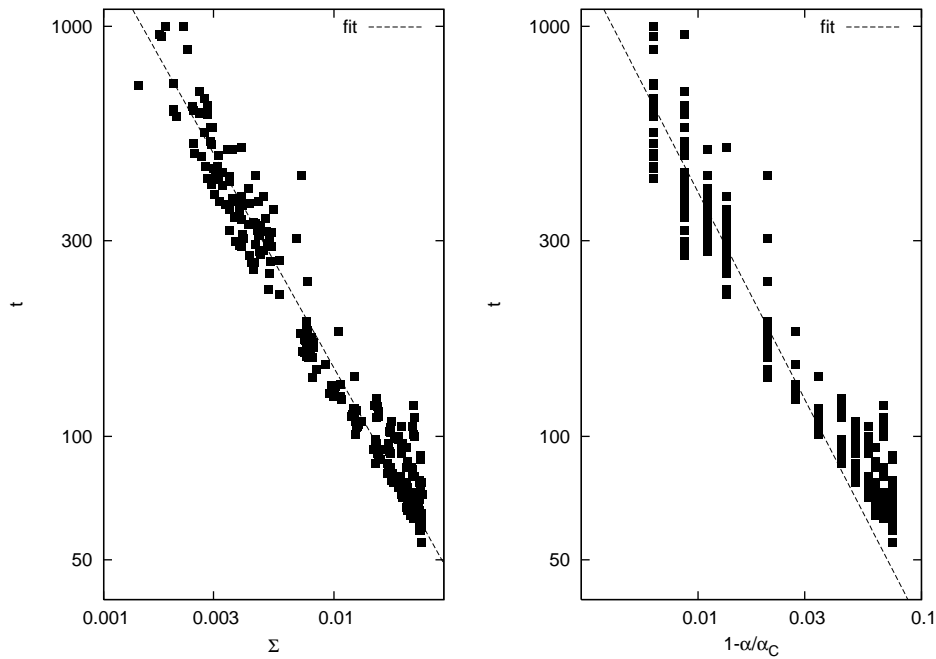}
\caption{Time of convergence as a function of $\Sigma$ (left) or $\alpha$ (right), using for $\pi$ the optimal intensity of the forcing ($N=10^5$). Close to the critical point, the time of convergence is shown to be inversely proportional either to $\Sigma$ or to $1-\frac{\alpha}{\alpha_C}$. The fit of the datas $t=t_0^\Sigma\times \Sigma^{-\beta}$ over $\Sigma\in[0,0.008]$ gives $t_0^\Sigma=1.474$ and $\beta=0.9994$. The fit $t=t_0^\alpha\times (1-\frac{\alpha}{\alpha_C})^{-\gamma}$ over $\alpha\in[4.14,4.25]$ gives $t_0^\alpha=3.02$ and $\gamma=1.057$.}\label{jo4}
\end{centering}
\end{figure}

We finally characterize the slowing down of the RA when approaching the
critical value $\alpha_C=4.2667$ (Fig.~\ref{jo4}). Far away from the critical
point, the time of convergence is highly reliable. Close to the critical
point, the algorithm gets slower and the time of
convergence becomes more variable. Setting the intensity of the forcing at its
optimal value, one  finds that, for $\alpha < \alpha_A$, the time of convergence is inversely
proportional to the complexity $\Sigma$~: $t=\frac{t_0^\Sigma}{\Sigma}$
($t_0^\Sigma=1.474$, Fig.~\ref{jo4}, left panel)~; moreover, as expected from
the pseudo-linear relationship between $\Sigma$ and $\alpha$, the time of
convergence also scales with $\frac{1}{\alpha-\alpha_C}$~:
$t=\frac{t_0^\alpha}{\alpha-\alpha_C}$ ($t_0^\alpha=3.02$, Fig.~\ref{jo4},
right panel).

Very close to $\alpha_C$ ($\alpha > \alpha_A$), the RA is unable to find ground states and only explores quasi-optimal assignments.

\section{Conclusion}\label{sect_concl}

This study  aimed at making fully distributed  the Survey Propagation decimation algorithm.

The first distributed solver, the ``SP diffusion algorithm'' diffuses as dynamical information the maximum bias over the system, so that variable-nodes can decide to freeze in a self-organized way. Its properties (solving rate, final number of active clauses when it reaches the paramagnetic state) are comparable with the previously described serialized ``SP Inspired Decimation''. The new feature is that the simulated time of convergence towards a solution, if this was implemented on a fully distributed device, goes as $\log(N)$, \textit{i.e.} scales optimally with the size of the graph.

The second solver, the ``SP reinforcement algorithm'', makes use of time-dependent local external forcings, which let the variables get completely polarized in the direction of a solution at the end of a single convergence. The estimated time of convergence of this solver, when implemented on a distributed device, also goes as $\log(N)$.

The present algorithm is the fastest existing solver of the K-SAT formulas, when implemented either on a distributed device \cite{chavas} or on a serialized computer. 

Moreover, the strategies proposed in the present manuscript are not specific to the K-SAT problem, but are likely to apply to many types of NP-complete constraint satisfaction problems (CSP) with  direct applications (e.g. data compression algorithm \cite{CMZ}).

\vspace{\baselineskip}

The different codes are available at (\cite{webcyril}, \cite{webserie}).

{\bf Acknowledgements}:
This work was supported by EVERGROW, integrated project No. 1935 in the
complex systems initiative of the Future and Emerging Technologies
directorate of the IST Priority, EU Sixth Framework. We thank Demian Battaglia and Alfredo Braunstein for very helpful discussions.

\end{document}